\documentclass[runningheads]{llncs}
\usepackage{graphicx}

\usepackage{graphicx}
\usepackage{multirow}
\usepackage{amsmath}
\usepackage{graphicx}
\usepackage{subcaption}
\usepackage{hyperref}
\usepackage{booktabs}
\usepackage{diagbox}
\usepackage[linesnumbered,ruled,vlined]{algorithm2e}
\usepackage{amssymb}
\usepackage{amssymb}
\usepackage{pifont}
\begin{document}
\title{Fetch-A-Set: A Large-Scale OCR-Free Benchmark for Historical Document Retrieval}
\titlerunning{Fetch-A-Set: OCR-Free HDR for Modern Times}
%
\author{Adrià Molina\inst{1}$^,$  \inst{2}\thanks{Main Corresponding Author}\orcidID{0000-0003-0167-8756} \and
Oriol Ramos Terrades\inst{1}$^,$  \inst{2}\orcidID{0000-0002-3333-8812} \and
Josep Lladós\inst{1}$^,$  \inst{2}\orcidID{0000-0002-4533-4739}}
\authorrunning{A. Molina et al.}
%
\institute{Computer Vision Center \and 
Computer Science Department \\
Universitat Autònoma de Barcelona, Catalunya \\
\email{\{amolina, oriolrt, josep\}@cvc.uab.cat} }

\maketitle              
\begin{abstract}
This paper introduces Fetch-A-Set (FAS), a comprehensive benchmark tailored for legislative historical document analysis systems, addressing the challenges of large-scale document retrieval in historical contexts. The benchmark comprises a vast repository of documents dating back to the XVII century, serving both as a training resource and an evaluation benchmark for retrieval systems. It fills a critical gap in the literature by focusing on complex extractive tasks within the domain of cultural heritage. The proposed benchmark tackles the multifaceted problem of historical document analysis, including text-to-image retrieval for queries and image-to-text topic extraction from document fragments, all while accommodating varying levels of document legibility. This benchmark aims to spur advancements in the field by providing baselines and data for the development and evaluation of robust historical document retrieval systems, particularly in scenarios characterized by wide historical spectrum.

\end{abstract}
\section{Introduction}
\label{sec:intro}
The automation of document understanding procedures is a growing trend across various industries. Document management systems to extract information, index, summarize or assist in decision making tasks are more and more frequent in fintech, legaltech, insurancetech, among other. A particular case are the systems for smart digitization of historical documents in archives and libraries. With historical data gaining significance in governmental bodies, heritage management is not exempt from this shift towards automation.

\begin{figure}[h!]
\centering
\includegraphics[width = \textwidth]{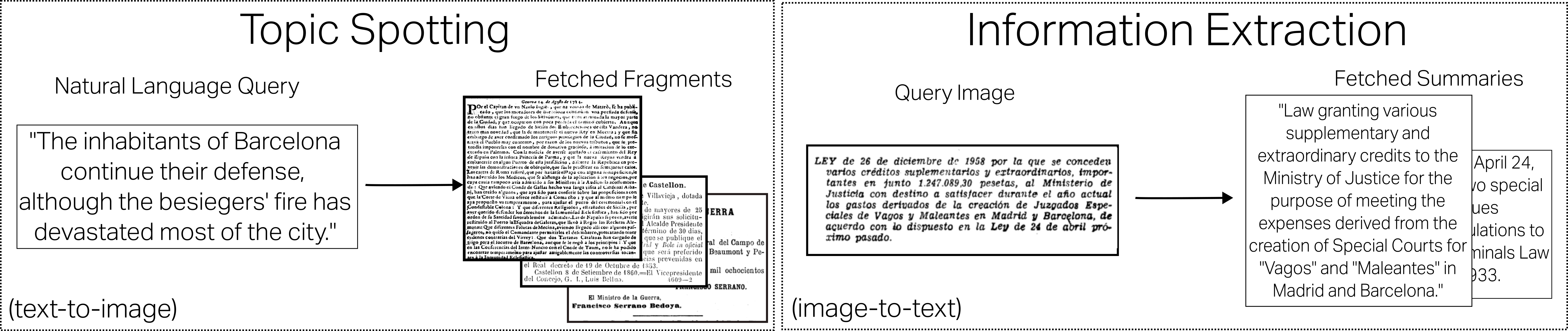}
\caption{Illustration depicting the primary objectives pursued by FAS in evaluating and training information extraction systems. Specifically, retrieving document fragments based on a given topic in natural language (right) or generating plausible descriptions from a given fragment (left).}
\label{fig:advertisement}
\end{figure}

With the growing of Document Intelligence, large scale digitization processes of historical documents for digital preservation purposes have scaled up to systems that are able to understand the contents providing innovative services to different communities of users. In this paper we consider two challenges faced by heritage institutions in handling vast documental sources. Firstly, from the user perspective, we tackle the need for continuous indexing of databases to enable natural language queries as a ``text-to-image" task. In other words, we aim to fetch relevant documents (image) based on human-written queries (text). We refer to this semantic-based task as \textit{topic spotting} (Figure \ref{fig:advertisement}, left), differentiating it from the tradictional word spotting that looks for exact matchings between the query words and the document content. 
Secondly, on the institutional front, managing large databases with millions of historical records becomes impractical for humans, causing delays in public access. Recognizing the importance of categorizing historical data in archival procedures, our paper also explores the ``image-to-text" task. This task, namely \textit{information extraction} (Figure \ref{fig:advertisement}, right), seeks to provide a feasible set of texts from an image, aiding users in establishing prior knowledge automatically for incoming data. Therefore, incorporating complex understanding tasks in the realm of historical document analysis systems will result in novel services to understand the history. 



The inherent nature of historical documentation renders the reliance solely on text-based approaches for information retrieval unfeasible \cite{hamdi2023depth}. The damaged condition of the documents, the lack of accurate ground truth, and the complexities of dealing with multiple languages severely impede the effectiveness of text-based methods. Consequently, it becomes crucial to incorporate visual insights to overcome these limitations and enable topic-sensitive retrieval in historical document analysis. Note that the existing historical document benchmarks \cite{nikolaidou2022survey} mainly concentrate on tasks such as classification \cite{jaume2019funsd}, date and writer retrieval \cite{christlein2019icdar}, or word spotting \cite{krishnan2019hwnet}, but they do not specifically address the challenges associated with topic-aware document retrieval. This scarcity of dedicated datasets underscores the need for research that explores document understanding in historical contexts.

This work proposes Fetch-A-Set (\textbf{FAS}) as a benchmark to evaluate the effectiveness of document understanding systems in fetching relevant information directly from natural text, eliminating the need for expensive OCR solutions, especially for large historical collections with significant temporal variance.

To demonstrate the relevance of this evaluation to the community, we summarize the contributions of this work as follows:
\begin{list}{$\circ$}{}  
    
    \item We introduce FAS as a benchmark for evaluating content-based extractive systems on historical records. FAS significantly enhances data availability, encompassing 400K samples sourced from three centuries of Spanish legislative documents (Section \ref{sec:data}).
    \item To demonstrate FAS's significance and its potential to inspire further research, we present a classical OCR-based baseline alongside a vision-based approach. We illustrate the contexts in which each method excels, providing guidance to authors for conducting more effective comparisons in their methodologies (Sections \ref{sec:arch}, \ref{sec:exp}).
    \item Finally, in Section \ref{sec:discussion}, we explore how historical bias may contribute to improve performance in low-legibility scenarios.
\end{list}

\section{Related Work}
\label{sec:sota}
As introduced in Section \ref{sec:intro}, the availability of OCR annotations in historical data is usually scarce. Moreover, recent analysis conducted by Hamdi \textit{et al.} (2023) \cite{hamdi2023depth} reveals that document extractive tasks achieve an accuracy of 80-90\% when applied directly to textual data benchmarks. However, in the same study, it is noted that integrating a visual perspective requires character recognition, which tends to drop the performance as the recognition systems are not adapted for such data. In historical document analysis, noise in OCR predictions poses a common challenge. In this context, retrieval tasks aim to directly index content from the visual domain. However, this section will explore how the historical document analysis community has predominantly focused on word spotting, driven by the necessity upon which this work is based: indexing text in large databases.

In fact, Nikolaidiou et al. (2022) \cite{nikolaidou2022survey} conducted an extensive survey on document analysis in historical contexts. Out of the 65 datasets they considered, 14 were identified as holding retrieval tasks. These tasks, detailed in Table \ref{tab:hist_doc}, primarily revolve around word spotting. Notably, 10 to 13 datasets focus on this task, with ``IAM-HistDB" often treated as a single entity consisting on \cite{fischer2011transcription} and \cite{fischer2012lexicon}. However, other historical document-related tasks emphasize visual properties over textual patterns. For instance, some tasks involve estimating dates from scanned photographs \cite{muller2017picture,net2024transformer,stacchio2020imago} or retrieving iconography/ornaments~ \cite{sovan2016public} from medieval documents. Additionally, certain tasks, like writer identification~\cite{christlein2019icdar,cilia2021papyrow,fiel2017icdar2017}, tackle subtle intra-class differences (visual attributes) alongside exact pattern processing (recognition).

Historical document retrieval literature commonly emphasizes textual content characterization (Table \ref{tab:hist_doc}), particularly through word spotting and writer identification tasks. This preference indicates a significant interest in deciphering the written content of documents.

\begin{table}[]
\begin{center}\resizebox{\textwidth}{!}{

\begin{tabular}{@{}lccccc@{}}
\toprule

Dataset \tiny{(HDR)} &
  Retrieval &
  \multicolumn{1}{c}{\#Samples \tiny{($\sim$)}} &
  \multicolumn{1}{c}{Time Span} & Type & Script \\ \midrule \midrule
  GERMANA\tiny{2009} \cite{perez2009germana}  & WS & 200k & 1891 & HW & Latin \\
  RODRIGO \tiny{2010} \cite{serrano2010rodrigo}  & WS & 200K & 1545 & HW & Latin \\
    Saint-Gall\tiny{2011} \cite{fischer2011transcription}  & WS & 5K & IX c. & HW & Latin \\
    Parzival\tiny{2012} \cite{fischer2012lexicon}  & WS & 20K & XIII c. & HW & Gothic \\
    Washington\tiny{2012} \cite{fischer2012lexicon}  & WS & 5K & XVIII c. & HW & Latin \\
    Esposalles\tiny{2013} \cite{romero2013esposalles}  & WS & 2K & 1451 - 1495 & HW & Latin \\
  BH2M\tiny{2014} \cite{fernandez2014bh2m}  & WS & 50K & 1617 - 1619 & HW & Latin \\
  HADARA80P\tiny{2014} \cite{panke2014historical} & WS & 15K & 1430 & HW & Arabic\\
  ENP\tiny{2015} \cite{clausner2015enp} & WS & - & XVIII - XX c. & Printed &Latin\\
  GRPOLY-DB\tiny{2015} \cite{gatos2015grpoly} & WS  & 100k & 1838 - 1912 & Mixed & Greek \\
   AMADI \tiny{2016} \cite{kesiman2016amadi} & WS & 10K & XV c. & Lontar &  Hanacaraka \\
    VML-HD\tiny{2017} \cite{kassis2017vmlhd}& WS & 200K & 1088-1451 & HW & Arabic\\
  CFRAMUZ \tiny{2017} \cite{arvanitopoulos2017cframuz} & WS & 20K & 1910-1946 & HW & Latin \\

 \midrule
  DocExplore \tiny{2016} \cite{sovan2016public} & Ornament & 2K  & X - XVI c. & Image & -\\
  \midrule

  ICDAR17 H-WI\tiny{2017} \cite{fiel2017icdar2017} & Writer & 5K & XIII - XX c.& HW & Latin \\
ICDAR19-HDRC-IR \tiny{2019} \cite{christlein2019icdar} & Writer & 20K & IX - XVII c.& HW & Latin \\

  Papy-Row \tiny{2021} \cite{cilia2021papyrow} & Writer & 6K & VI c. & Papyrus & Greek \\
  
\midrule
 HistDIA \tiny{2021} \cite{seuret2021icdar} & Date, Font, Loc & 10/30/5 K & IX-XVIII c. & HW & Latin \\
\midrule
DEW \tiny{2017} \cite{muller2017picture} & Date & 1M & 1930 - 1999 & Image & -\\
IMAGO \tiny{2020} \cite{stacchio2020imago} & Date & 80K & 1845 - 2009 & Image & -\\
DEW-B \tiny{2024} \cite{net2024transformer} & Date & 1.5M & 1930 - 1999 & Image & -\\

 \bottomrule
\end{tabular}}
\end{center}
\caption{Table for most popular historical document retrieval datasets, "retrieval" stands for the retrieved content in the task: Word Spotting (\textbf{WS}), \textbf{Ornament}, \textbf{Writer} and \textbf{Font} Identification and \textbf{Date} and \textbf{Loc}ation estimation.}
\label{tab:hist_doc}

\end{table} 

Consequently, the upcoming sections will introduce the \textbf{FAS} benchmark. This dataset serves as an endeavor to assess systems designed to extract meaning from historical documents, moving beyond mere word spotting and delving into the comprehension of textual content from the vision perspective.

\section{The Fetch-A-Set Dataset}
\label{sec:data}
In this section, the construction and basic analytics of the \textbf{FAS} dataset are presented. This benchmark comprises a set of full-page documents with an associated natural language query. We design a solid heuristic for assigning a one-to-one relationship between fragments of the document and queries. In this way, we expect retrieval systems to be able to replicate this correspondences out of a randomly selected distractor set.
\begin{figure}[!tbp]
    \centering
    \includegraphics[width=0.85\textwidth]{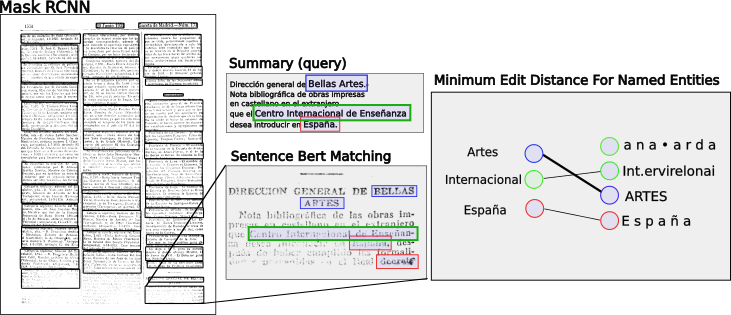}

    \caption{Query-Region Matching system for creating the Ground Truth (GT) for FAS dataset. Since the presence of the fragment $F$ in the document $d$ is human-annotated, and the number of invalid fragments within a page is typically low, the risk of adding noise is significantly reduced.}
    \label{fig:scheme}
\end{figure}

\subsection{Building the FAS dataset}

Our research extensively utilizes the publicly accessible repository of the ``Boletín Oficial del Estado" (BOE), Spain's Official State Gazette, a pivotal source for disseminating government-approved laws, decrees, provisions, resolutions, and regulations \footnote{https://www.boe.es/buscar/gazeta.php}. This repository, accessible online, encompasses historical documents dating back to the XVII century, containing approximately one million pairs of human annotated \textbf{summaries} ($q$) and corresponding \textbf{documents} ($d$), along with OCR text extracted with a commercial recognition system \footnote{\texttt{Abbyy Recognition Server v4}}. As expected and discussed in Section \ref{sec:intro}, the transcriptions are imperfect. The sentence-bert distance \cite{reimers2019sentencebert} between the \textbf{OCR transcription} and the queries ($q$) is considered as measurement for \textbf{legibility}; as noisy transcriptions tend gradually lose meaningfulness with respect the original text as the transcription degrades.

\begin{figure}[]
    \centering
    \includegraphics[width=0.9\textwidth]{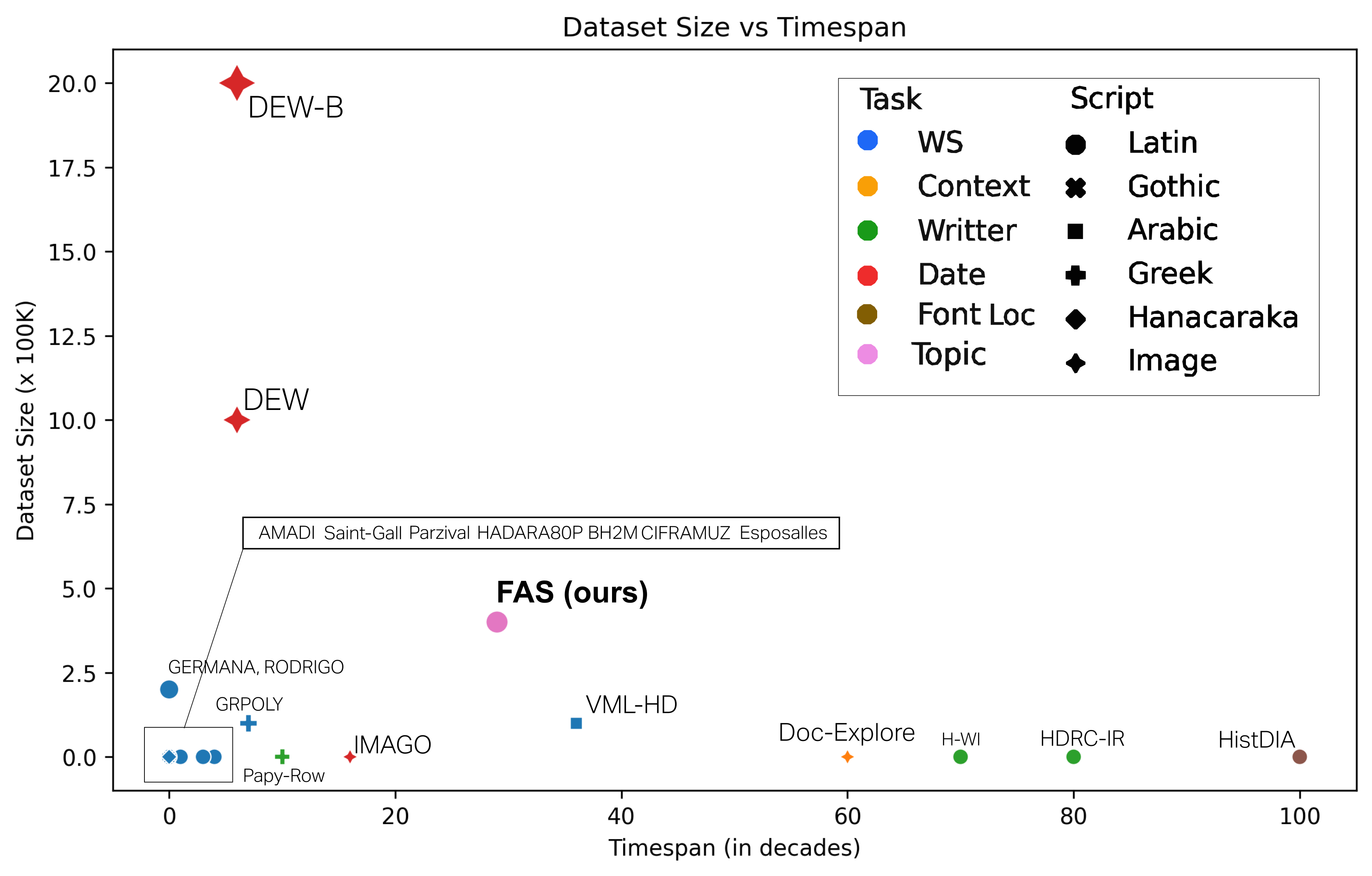}

    \caption{A scatterplot depicting FAS as the preeminent historical document-based retrieval benchmark among those previously examined, while also maintaining a substantial temporal scope. This characteristic enhances the robustness of historical document analysis systems by encompassing wider temporal breadths.}
    \label{fig:resume}
\end{figure}

Given the multifaceted nature of newspapers, meaning that (specially in most modern samples) a document contains some irrelevant fragments on the page for a few relevant ones, one of the central challenges in curating this dataset is associating queries $q$ with specific fragments $F$ within documents. To address this, we implement a two-step selection process: first, a Mask-RCNN \cite{he2017mask} model trained on the Prima Layout Analysis Dataset \cite{antonacopoulos2009realistic} is used to identify relevant document regions, followed by matching queries with OCR text using sentence-bert encoding \cite{reimers2019sentencebert}. Notably, the OCR transcriptions may be unreliable, and to enhance the accuracy of query-document associations, a filtering step employing the Hungarian algorithm \cite{kuhn1955hungarian} with edit distance measure is used to assess the similarity of named entities\footnote{Extracted with SpaCy \texttt{https://spacy.io/models/es}. } as illustrated in Figure \ref{fig:scheme}.

In adopting this approach,we have established a robust foundation for ensuring the accuracy of information extraction (\textbf{image-to-text}) and topic spotting (\textbf{text-to-image}). To manage practical and computational constraints, we initially focused on a well-curated subset of 400K \textbf{fragments of documents} \footnote{Train / Test / Distractor splits with queries, hyper-references to the documents (whole page), regions of interest and OCR transcriptions can be found at \texttt{http://datasets.cvc.uab.es/BOEv2/BOEv2.zip}.}. As seen in \ref{tab:splits}, we randomly divide the given set of documents in a \textbf{train} and \textbf{test} split, with an additional set of 1024 \textbf{distractor} documents that will serve as an evaluation anchor for retrieval in order to avoid expensive computations and the usage of the whole training set as a distractor, which, due to the challenges of the task, would drop the performance to non-significant comparisons.

\begin{table}[]
    \centering{
        
        \begin{tabular}{@{}rrr@{}}
        \toprule
        \#Train & \#Test & \#Distractors \\ \midrule
        384567 & 42997 & 1024 \\ \bottomrule
        \end{tabular}

}
\vspace{1px}
\caption{Table showing train, test and distractor number of samples in FAS.}
\label{tab:splits}
\end{table}

\subsection{Dataset Analytics}

As illustrated in Figure \ref{fig:resume} and detailed in Table \ref{tab:hist_doc}, the majority of retrieval datasets primarily emphasize word spotting tasks. However, it is worth noting a significant bias towards datasets with narrower time spans, with the exception of VML-HD \cite{kassis2017vmlhd}. Despite this bias, certain benchmarks focusing on textual analysis, such as \cite{christlein2019icdar,fiel2017icdar2017,seuret2021icdar}, extend the temporal range up to 1000 years.

\begin{figure}
  \centering
    \includegraphics[width=\textwidth]{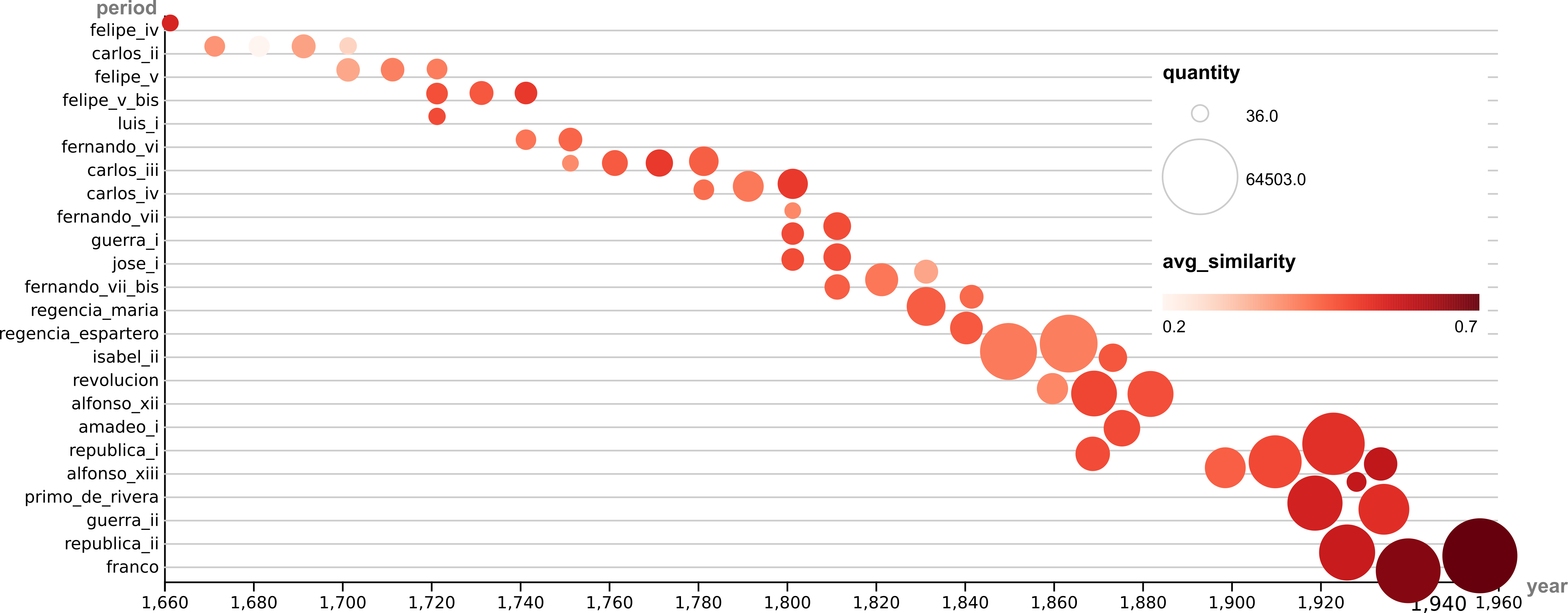}

  \caption{Beeswarm plot showing the time arrow of the dataset with the years (X-Axis), historical period (Y-Axis) the quantity of documents (size) and the average legibility score (hue).}
  \label{fig:swarm}
\end{figure}

While the pursuit of temporal diversity may seem like a matter of mere curiosity to the general audience, in certain contexts, it holds significant relevance. Specifically, the consideration of systems that continuously ingest historical documents from various sources can offer valuable insights. In the realm of historical document analysis, methods often prioritize robustness by focusing narrowly on specific applications (see Figure \ref{fig:resume}), neglecting the potential benefits of incorporating temporal diversity. In this regard, the FAS framework emerges as a pivotal convergence point between variance and scale. It addresses the need for extractive textual tasks in retrieval systems, positioned at the top of scale in text analysis while encompassing a significant time span.

\begin{figure}
  \centering
    \includegraphics[width=0.48\textwidth]{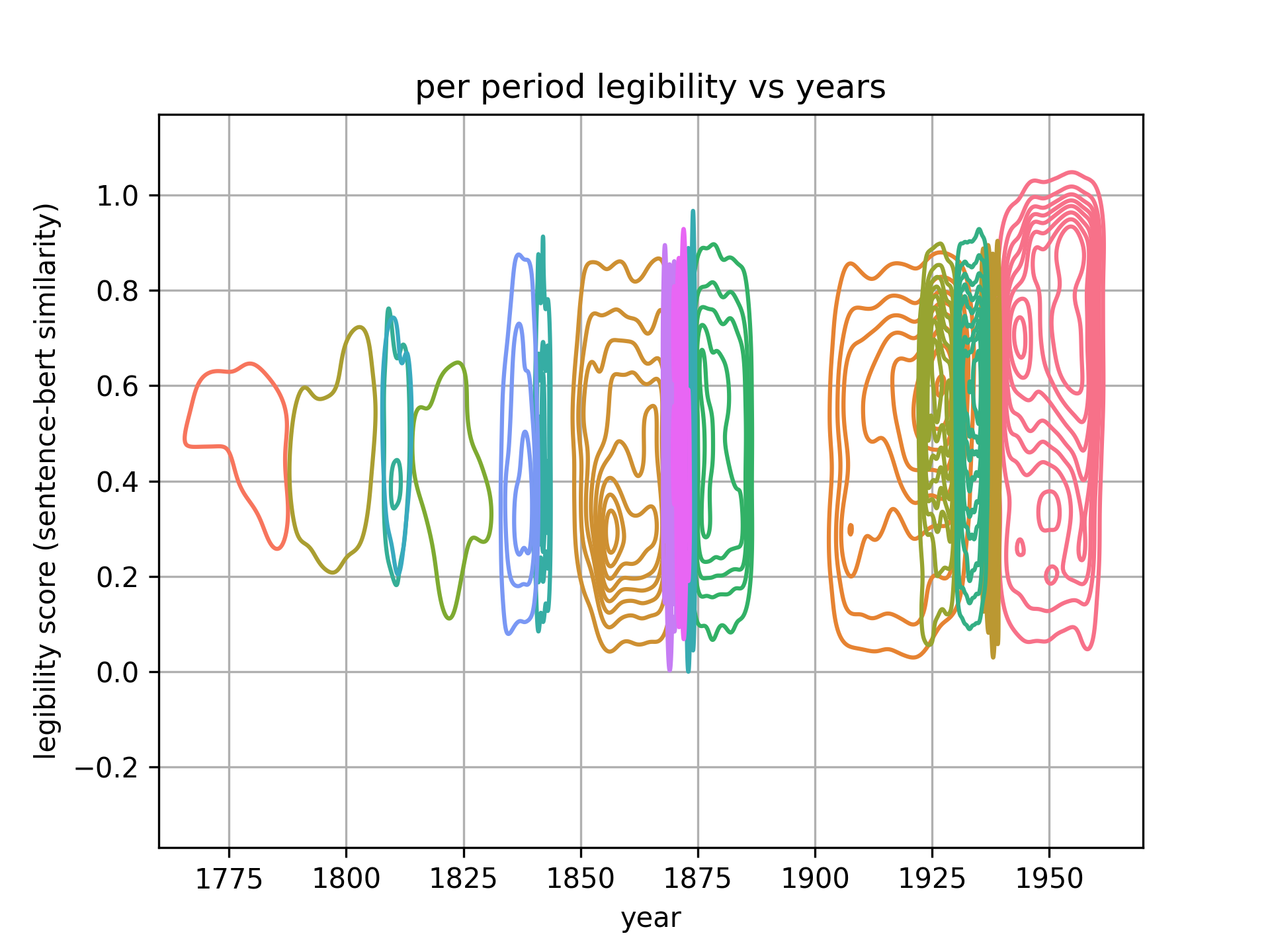}
    \includegraphics[width=0.48\textwidth]{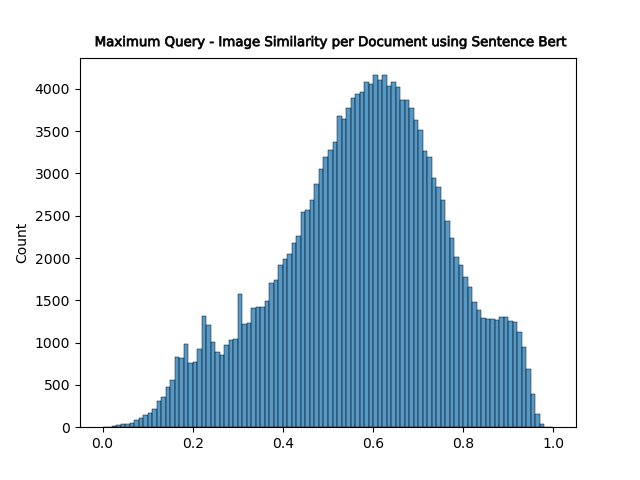}

  \caption{Left: Distribution of legibility (Y-Axis) through the years (X-Axis) per historical period (hue). Right: Global distribution of legibility score.}
  \label{fig:kde}
\end{figure}

As shown in Figure \ref{fig:swarm}, around 100k samples from the FAS dataset belong to the period of time of Franco's fascist dictatorship over Spain (1939-1975, limited to 1959 by BOE's organization); followed (80K samples) by Alfonso XIII's reign (1902-1923 / 1931). This unbalance, as a consequence of both time variance of periods and increasing contemporaneity of the events, is furtherly explored in Figure \ref{fig:kde}; where we observe a clearer correlation between legibility and years. 

As previously noted in Figure \ref{fig:swarm}, it seems that average legibility is expressed through the years. While it may become apparent that legibility increases with time, it is an effect of averaging scores. In Figure \ref{fig:kde} we conclude that high legibility has a bias towards modernity, but not the other way around, as many samples in most modern documents also contain poor legibility scores. As variance in deployed document retrieval systems is a mandatory property when pursuing robustness, this captured variance in most modern documents shall promote dataset's strength in contributing to the historical document analysis community.

\begin{figure}[]
\resizebox{\textwidth}{!}{

    \begin{minipage}{0.48\textwidth}
            \centering\includegraphics[width=1\textwidth]{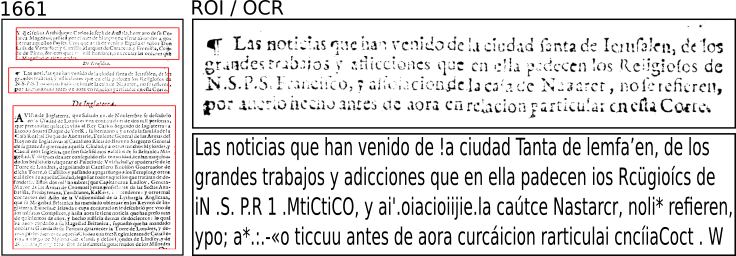}
            \centering\includegraphics[width=1\textwidth]{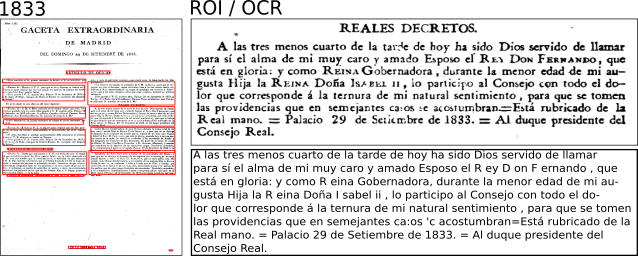}
    \end{minipage}
    \hspace{0.005\textwidth}
    \begin{minipage}{0.48\textwidth}
        \centering\includegraphics[width=.9\textwidth]{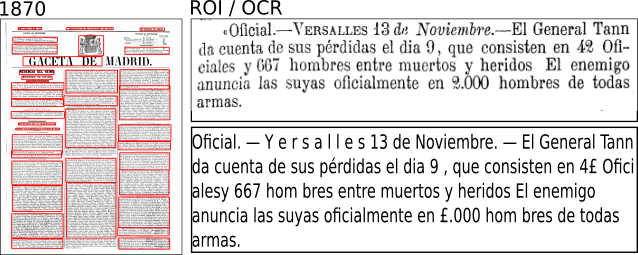}
        \centering\includegraphics[width=.9\textwidth]{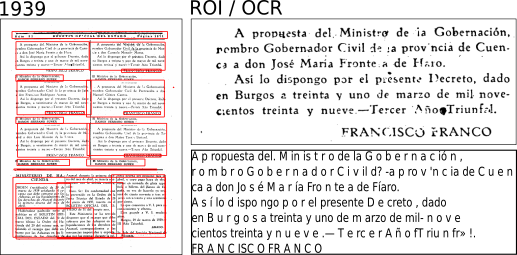}
    \end{minipage}}
\caption{Examples of segmented documents (1661, 1833, 1870 and 1939) among its corresponding ROI and extracted OCR.}
\label{fig:example}
\end{figure}

As depicted in Figure \ref{fig:example}, the extensive temporal breadth results in considerable variability in layout. This variability holds promise for the development of systems intended to accurately capture such contextual variance in their application domain.


\section{Baselines}
\label{sec:arch}
As previously introduced, the primary objective of this study is to tackle the challenge of evaluating information extraction systems within a retrieval framework. We formulate a multi-modal retrieval problem that can be perceived as both information extraction (image-to-text) or topic spotting (text-to-image). In this section, we propose two different baselines serving as pivot to measure the performance of furhterly developed retrieval systems. To achieve this, we leverage two distinct views for each document $d$: firstly, the query $q$, which is encoded using a text encoder $\tau = \phi_t(q)$, and secondly, the image representing the corresponding fragment $F$, encoded through a visual embedding $\nu = \phi_v(F)$. These two distinct representations—textual and visual—allow us to explore information retrieval by bridging the gap between images and natural language queries. Since newspapers usually contain a wide variety of topics for each page, the fragment $F$ is the region of the document which corresponds to the natural language description $q$. 

\begin{figure}
    \centering
    \includegraphics[width=\textwidth]{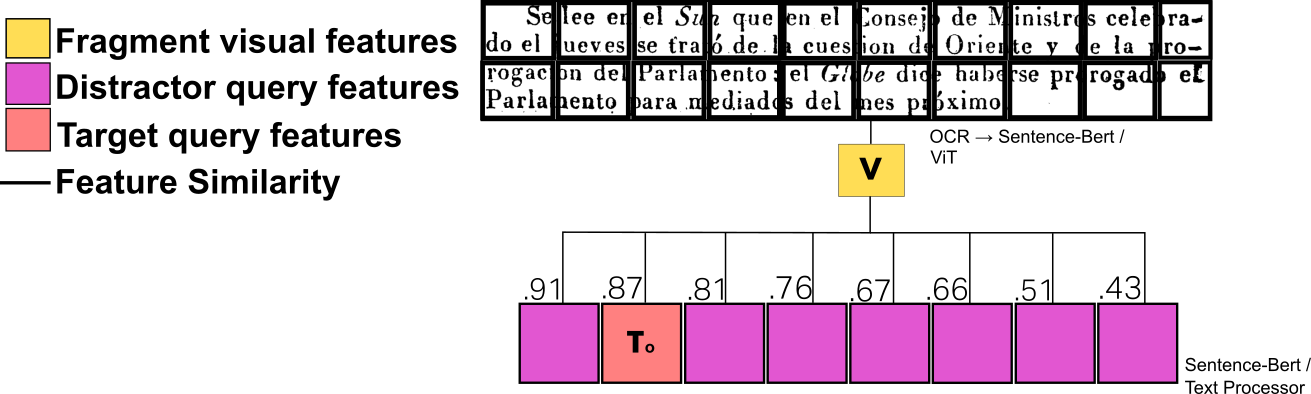}
    \caption{Basic scheme of the proposed methods where a set of features $\nu$ are extracted from a fragment. Then, the corresponding query feature $\tau_0$ is inserted among the set of distractor queries. When comparing distances, the tuple ($\nu$, $\tau_0$) is expected to show maximum similarity. For text-to-image task, the objective is analogous.}
    \label{fig:method}
\end{figure}

In order to show the relevance of this benchmark to authors and practitioners, two baselines are defined below to test the evaluation capabilities of FAS. As shown in Figure \ref{fig:method}, both options are retrieval-based. In the case of the vision-based approach, a fragment is encoded at pixel-level to obtain a close representation to the query. In the case of the OCR-Based approach, we take advantage of the available OCR in the dataset to get an encoded representation of both query and OCR texts. 

\textbf{Vision-Based Approach}
In order to obtain a robust representation of the content at pixel level, we propose the usage of a Vision Transformer \cite{dosovitskiy2020image} (ViT-B/32) with 32x32 patch size pretrained on CLIP \cite{radford2021learning}. This ViT embeddings are linearly projected and fine-tuned into the \textit{topic space}, which is shared with the query encoder. In doing so, we make use of the standard transformer architecture \cite{shazeer2020glu,vaswani2017attention,xiong2020layer,zhang2019root} with 2 layers and 4 heads. We take the decision of training the query embeddings from scratch since the complexity of the textual information present in the dataset does not appear to hold the necessity of bigger architectures. We simultaneously fine-tune the visual encoder and train the query space to be projected at close points in the \textit{topic space}. For doing so, we leverage triplets for using metric learning approaches such as Triplet Margin \cite{balntas2016learning}, or contrastive losses such as SimCLR \cite{chen2020simple,oord2018representation} and CLIP \cite{radford2021learning}. In this scenario, the fragment representation $v$ is encoded through a fine-tuned ViT, while the query $q$ is processed through a text encoder to obtain a representation $\tau$.

\textbf{OCR-Based Approach}
In order to conduct a comparison with a state-of-the-art text-based approach, we take advantage of the OCR transcriptions acquired with the commercial OCR present in the dataset. We consider the document features as the recognized text processed through the same \textit{sentence-bert}\cite{reimers2019sentencebert} encoder than the queries. We then explore the extend to which a given document matches its corresponding query whenever its distance is minimum among the distractor set.

\section{Experiments}
\label{sec:exp}
\subsection{Objective}
\label{sec:objects}
The evaluation proposed in the \textbf{FAS} benchmark, aims to minimize the distance between the encoded representations of the target document or query ($\tau$, $\nu$) and all elements within the distractor set ($\mathfrak{Q}$). Mathematically, the objective (image-to-text) is expressed as $\text{min} \:  \lVert \phi_t(q_o) - \phi_v(F_o) \rVert$ and, $\forall q_i \in \mathfrak{Q}$, $\text{max} \: \lVert \phi_t(q_i) - \phi_v(F_o) \rVert $ with $\{q_o\} \cap \mathfrak{Q} = \O$. This objective aims to identify the query ($q$) within the given distractor set ($\mathfrak{Q}$) that exhibits the minimum distance to the encoded representation of the target fragment ($F_o$), being $q_o$ the correct correspondence. The encoded representations, derived from the text ($\phi_t$) and visual ($\phi_v$) embeddings, enable the retrieval system to effectively match queries and documents within a multi-modal retrieval framework.

\subsection{Evaluation Metrics}
The previously defined objective, shall be evaluated using top-K Accuracy \textbf{Acc@\{1, 5, 10\}} and, in some experiments, average ranking (\textbf{AR}) indicating the average position (0, $|\mathfrak{Q}|$) where the correct matching ($\tau$ or $\nu$) has been placed with respect the set of distractor data $\mathfrak{Q}$.
\subsection{Results}
\begin{table}[t]
\begin{center}

\begin{tabular}{@{}cccrrrrrr@{}}
\toprule
\multicolumn{2}{l}{Approach}                           &  & \multicolumn{5}{c}{Topic Spotting (text-to-image) Metrics} & \\ \midrule
Architecture & Loss &  & \multicolumn{1}{l}{Acc@1} & Acc@5& \multicolumn{1}{l}{Acc@10} & \multicolumn{1}{l}{AvgRank $\downarrow$} & \multicolumn{1}{l}{Acc@$1_{\text{low}}$} & \multicolumn{1}{l}{Acc@$5_{\text{low}}$} \\ \cmidrule(r){1-2} \cmidrule(l){4-9} 
\multirow{4}{*}{\parbox{2cm}{ViT-B/32 \\\centering \tiny{\cite{radford2021learning,dosovitskiy2020image}}}}& Triplet \cite{balntas2016learning}&  & 0.473&0.639& 0.695& 52.74 & 0.289& 0.419\\
                               & SimCLR\cite{chen2020simple}                 &  & 0.509&0.672& 0.723& 44.12& \textbf{0.296}& \textbf{0.443}\\
                               & CLIP\cite{radford2021learning}                  &  & 0.460&0.657& 0.719& 32.33 & 0.264& \textbf{0.447}\\ \midrule
\multicolumn{1}{l}{OCR + sBert\cite{reimers2019sentencebert}} & \multicolumn{1}{c}{-} &  & \textbf{0.657}&\textbf{0.785}& \textbf{0.837}& \textbf{16.46}& 0.124& 0.304\\  \bottomrule
\end{tabular}

\vspace{3pt}

\begin{tabular}{@{}cccrrrrrr@{}}
\toprule
\multicolumn{2}{l}{Approach}                           &  & \multicolumn{5}{c}{I. Extraction (image-to-text) Metrics} & \\ \midrule
Architecture & Loss &  & \multicolumn{1}{l}{Acc@1} & Acc@5& \multicolumn{1}{l}{Acc@10} & \multicolumn{1}{l}{AvgRank $\downarrow$} & \multicolumn{1}{l}{Acc@$1_{\text{low}}$} & \multicolumn{1}{l}{Acc@$5_{\text{low}}$} \\ \cmidrule(r){1-2} \cmidrule(l){4-9} 
\multirow{4}{*}{\parbox{2cm}{ViT-B/32 \\\centering \tiny{\cite{radford2021learning,dosovitskiy2020image}}}}       & Triplet\cite{balntas2016learning}&  &0.523&0.640& 0.692& 64.00& 0.326& 0.421\\
                               & SimCLR\cite{chen2020simple}                 &  & 0.522&0.675& 0.726&  43.05& \textbf{0.341}& 0.458\\
                               & CLIP\cite{radford2021learning}                  &  & 0.482&0.662& 0.725& \textbf{32.99}& 0.326& \textbf{0.464}\\ \midrule
\multicolumn{1}{l}{OCR + sBert\cite{reimers2019sentencebert}} & \multicolumn{1}{c}{-} &  & \textbf{0.623}&\textbf{0.723}&  \textbf{0.760}& 44.19& 0.073& 0.149\\  \bottomrule
\end{tabular}
\vspace{.25cm}
\caption{Text-based and Vision-Based baselines for topic spotting (top) and information extraction (bottom). In further analysis we show how visual baseline can outperform OCR+Bert in some challenging scenarios for recognition. The subindex \textit{low} stands for the percentile 0-25\% of worst legible documents.}

\label{tab:minitan}
\end{center}
\end{table}

In Table \ref{tab:minitan}, the quantitative results of both vision and OCR-based approaches are presented. As reported in \cite{musgrave2020metric}, it is expected to all metric learning approaches to perform competitively when compared with the other ones, which is the case for the evaluated visual baseline system.

Upon observation, we note that vision approaches prove quantitatively more advantageous than the OCR-based baseline in the context of tasks containing poorly legible documents. On the other hand, using text features becomes significantly more convenient in situations where the recognition can be performed correctly. 

The key takeaway from both textual and vision-based baselines is that text features perform exceptionally well when the text is clear and legible. However, there's potential for vision-based systems in scenarios where visual insights offer an advantage. As discussed in Section \ref{sec:discussion}, the visual benchmark incorporates both keyword spotting and layout-driven date estimation. This implies that hybrid systems capable of leveraging visual features when text is unavailable, and text features otherwise, could enhance the robustness of systems evaluated with this dataset.

\section{Discussion}
\label{sec:discussion}
From a qualitative standpoint, our primary emphasis is on evaluating both retrieval performance and the comprehensibility of the results. Given the consideration that, as shown in Section \ref{sec:exp} Table \ref{tab:minitan}, ViT performance seems to hold even in situations where legibility is low. In this section, we aim to explore which features are being utilized in terms of encoding topic information.
\begin{figure}
    \centering
    
    \includegraphics[width=\textwidth]{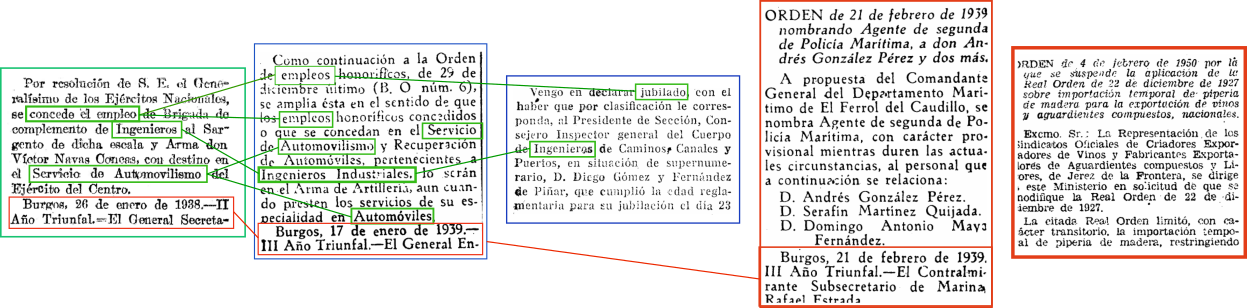}
    
\caption{Success case of the retrieval of documents related to the query \textit{``Promotion to the rank of Sergeant in the Complementary Engineer Brigade is hereby granted to Mr. Víctor Navas Concas"} (translated).  Some relevant semantic features (green), some signs of bias (red) and the relevant document ranked at the top (left, first).}
\label{fig:good-result-biased}

\end{figure}

In Figure \ref{fig:good-result-biased}, we observe a document that has been accurately ranked as the top result by our 1-Nearest Neighbor approach. Interestingly, the top-ranked documents contain significant content related to engineering job status. The presence of common keywords serves as a necessary, yet not sufficient, condition for the encoder to perform implicit recognition. However, there are also indications of potential bias in our results. All documents belong to the same period of time, some documents exhibit a shared historical fragment at the bottom (``II Victorious Year", ``III Victorious Year"), reminiscent of the early years of the fascist military uprising in Spain (1936-1939), therefore, it is important to exercise caution when interpreting the encoder's ability to recognize important words in the text.  

\begin{figure}
    \centering
    \includegraphics[width=0.3\textwidth]{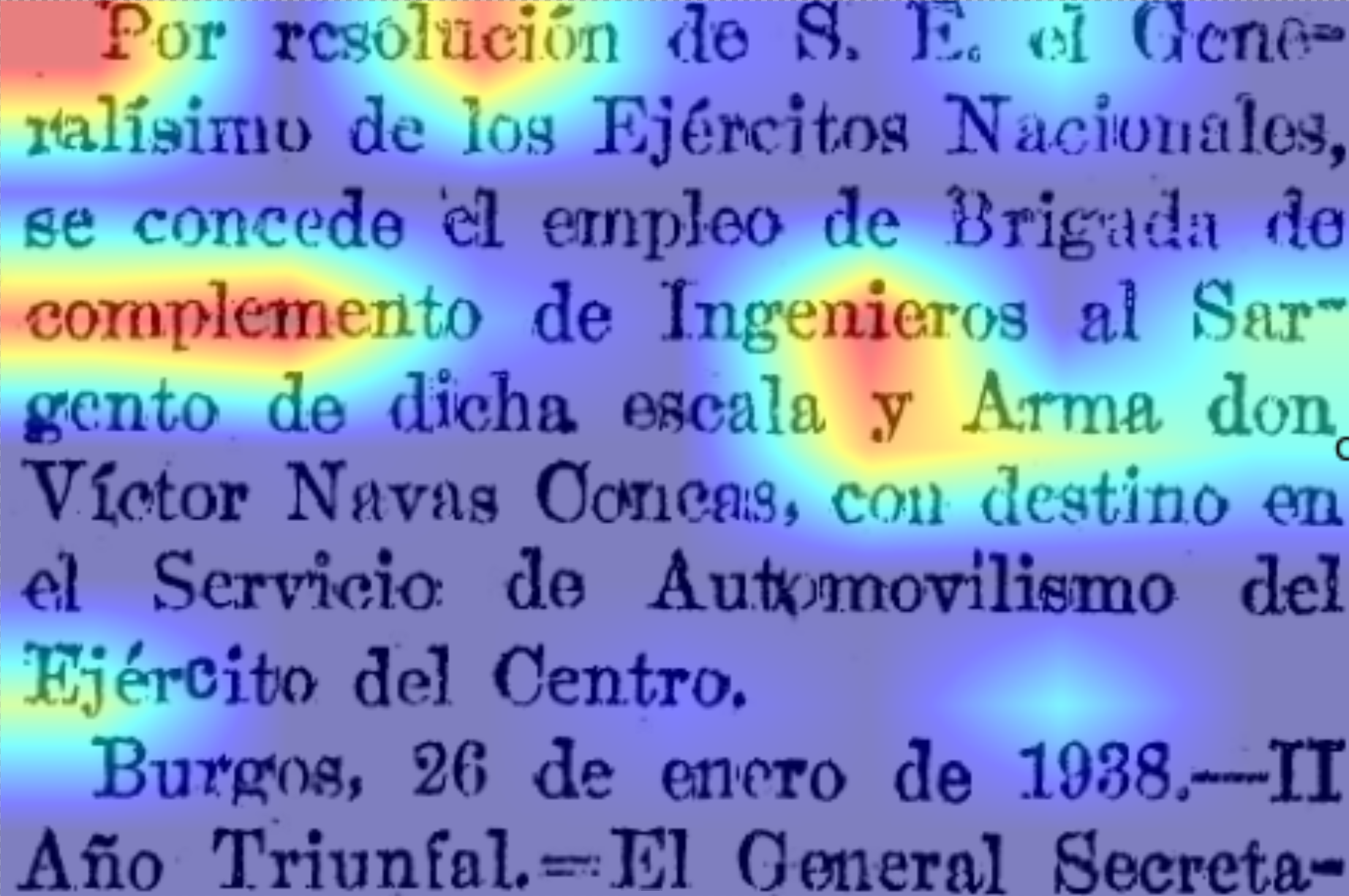}
    \includegraphics[width=0.50\textwidth]{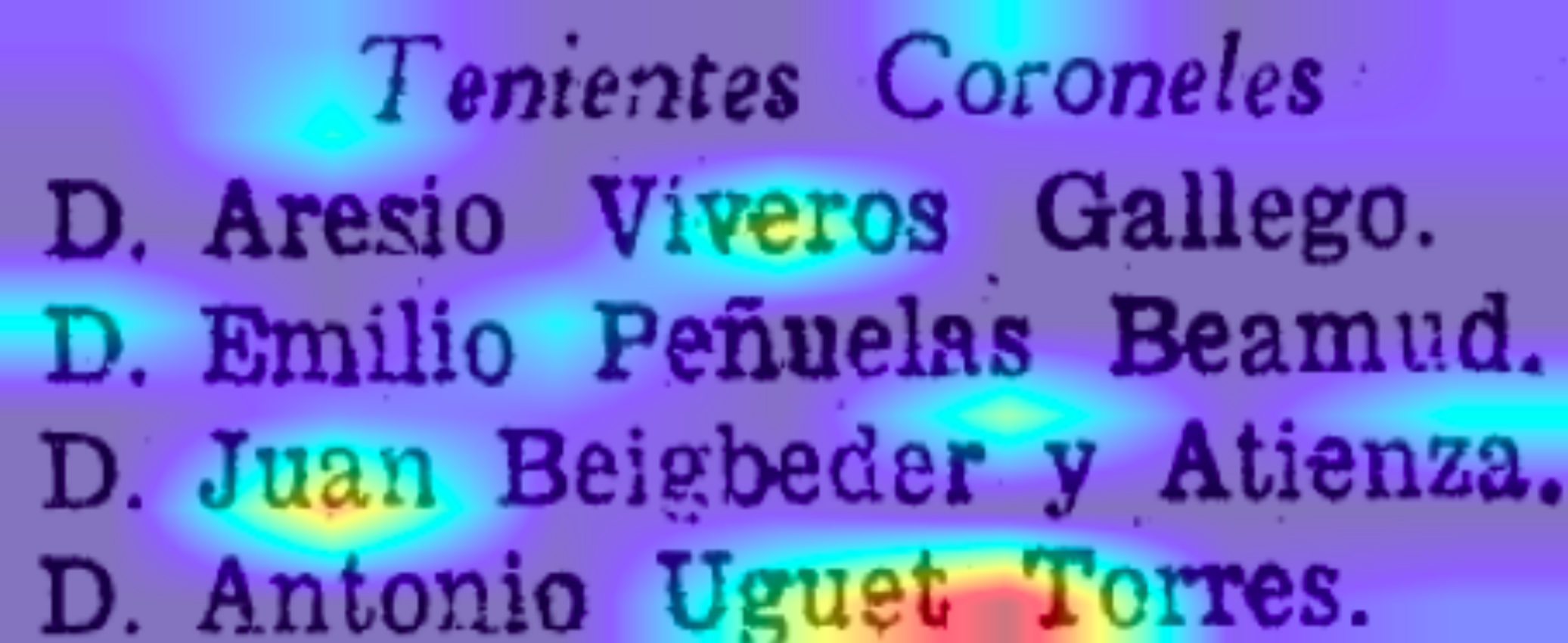}
    \caption{GradCam \cite{selvaraju2017grad} activations on fragments $F$ at the visual tokenizer of the ViT baseline when optimized towards $q$ alignment.}
    \label{fig:fragment_gradcam}
\end{figure}

\begin{figure}
    \centering
    \includegraphics[width=0.7\textwidth]{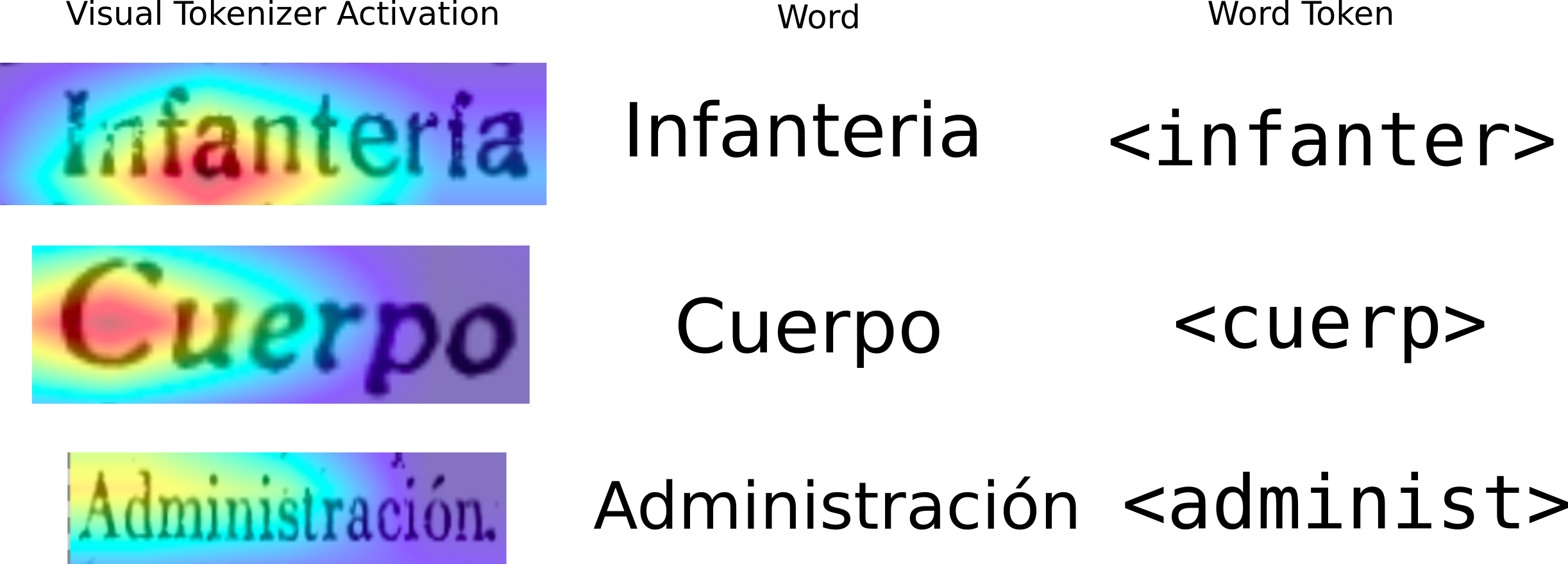}
    \caption{Examples of visual words (left) where the visual tokenizer (activations) prioritizes only a patch of the image rather than combining several patches.}
    \label{fig:words}
\end{figure}

To investigate whether the visual extractor relies on the text content or if other biases contribute to its performance, we conduct an inspection of the activations (\cite{selvaraju2017grad}) of the visual tokenizer given its ground truth topic (query, $q$) Figure \ref{fig:fragment_gradcam}. As it can be observed, there is no sign of bias towards the usage of the bottom fragment of the text, moreover, in Figure \ref{fig:words} it is shown how the visual representation of the fragment $F$ efficiently tokenizer words to its stems; which should be a sign of robustness to noise and degradation.

\begin{figure}[t]
    \begin{minipage}[t]{0.5\textwidth}
        \centering
        \begin{tabular}{@{}lcccccc@{}}
        \toprule
        \diagbox[width=6em]{Layer}{Augm\footnote{NxN: Average blurring $K_{size} = N$.}} & None & 2x2 & 4x4 & 8x8 & 16x16 & Flip \footnote{Flip: Horizontal and vertical flipping}\\ \midrule
        Tok & 71.8 & 79.8 & 85.4 & 88.5 & 118.2 & 70.3 \\
        LN1 & 63.2 & 70.2 & 121.5 & 79.9 & 94.3 & 64.2 \\
        $\text{Tr}_1$ & 36.8 & 37.4 & 36.4 & 32.9 & 35.6 & 37.4 \\
        $\text{Tr}_6$ & 17.2 & 18.7 & 21.7 & 22.1 & 24.6 & 21.5 \\
        $\text{Tr}_{12}$ & 14.4 & 16.1 & 19.8 & 22.1 & 26.1 & 23.2 \\
        LN2 & 11.6 & 12.8 & 15.6 & 17.9 & 21.6 & 18.2 \\
        Out & 17.2 & 16.0 & 19.2 & 20.1 & 24.2 & 20.9 \\ \bottomrule
        \end{tabular}
    \end{minipage}%
    \hspace{0.075\textwidth}
    \begin{minipage}[t]{0.4\textwidth}
        \centering
        \begin{minipage}[t]{0.4\textwidth}
            \centering
            \includegraphics[width=\textwidth]{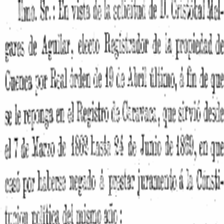}
        \end{minipage}\hfill
        \begin{minipage}[t]{0.4\textwidth}
            \centering
            \includegraphics[width=\textwidth]{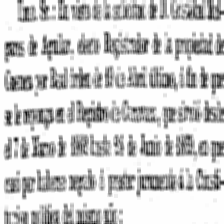}
        \end{minipage}\\[1em]
        \begin{minipage}[t]{0.4\textwidth}
            \centering
            \includegraphics[width=\textwidth]{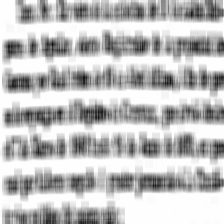}
        \end{minipage}\hfill
        \begin{minipage}[t]{0.4\textwidth}
            \centering
            \includegraphics[width=\textwidth]{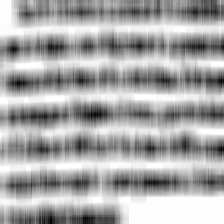}
        \end{minipage}
    \end{minipage}
        \caption{Mean Absolute Error on date estimation given the internal ViT representation of documents (tokenizer, normalization, transformer and output) given different sizes of blurring and examples on incrementally blurred documents.}
    \label{tab:date_estimation_table}
\end{figure}

However, it is apparent (Figure \ref{fig:good-result-biased}) than the ViT is incorporating some (benefitial) temporal bias; in Figure \ref{tab:date_estimation_table} we ask ourself whether this is happening at text level (which would contradict Figure \ref{fig:fragment_gradcam}) or ViT is focusing on some other features to incorporate temporal scale in the process of assigning a topic representation to a given image. 

As illustrated in Figure \ref{tab:date_estimation_table}, it is evident that the deeper the visual representation utilized, the more accurate the date estimation by a linear regression. This effect is substantial, with average errors ranging from 11 to 15 years in some cases. Moreover, this trend persists even with increasingly heavier degradations of the fragment, indicating the sustained performance of the model.

From these findings, we can infer that there is indeed temporal information embedded within the tokens that define the topic space (as depicted in Figure \ref{fig:good-result-biased}). However, while there is also sensitivity to text (as shown in Figures \ref{fig:fragment_gradcam} and \ref{fig:words}), the date estimation primarily occurs at the layout level rather than by associating specific words with the temporal scale (Figure \ref{tab:date_estimation_table}).

\section{Conclusions}
\label{sec:conc}
This paper has presented a novel dataset and baselines for evaluating historical document retrieval systems, termed Fetch-A-Set (FAS), which proves effective in assessing the performance of both text and vision-based systems. We underscore the necessity of transitioning from individual words to natural text to enhance historical document management in terms of indexing and comprehension, and we provide authors with a comprehensive dataset for training and evaluation purposes.

Furthermore, the paper stresses the importance of employing large-scale datasets for training and evaluation, as they encompass a broader temporal variance relevant to the application of systems developed by the community.

Lastly, we evaluate both baselines and investigate the impact of vision on the results. Our findings suggest that incorporating temporal bias into retrieval systems is advantageous, even without relying on text features. This exploration suggests that both vision and text are niche solutions for solving FAS, which should help to evaluate text, vision-based systems and hybridisation systems that aim to provide robustness by incorporating both modalities into the system.
\section*{Acknowledgment}

This work has been partially supported by the Spanish project PID2021-126808OB-I00, Ministerio de Ciencia e Innovación, the Departament de Cultura of the Generalitat de Catalunya, and the CERCA Program / Generalitat de Catalunya. Adrià Molina is funded with the PRE2022-101575 grant provided by MCIN / AEI / 10.13039 / 501100011033 and by the European Social Fund (FSE+).

\bibliographystyle{splncs04}
\bibliography{bibliography}

\end{document}